\newcommand{\pcmq}{\mbox{cm$^{-2}$}}
\newcommand{\psec}{\mbox{s$^{-1}$}}

\newcommand{\funit}{\mbox{ph~\pcmq~\psec}}

\newcommand{\eunit}{\mbox{erg~\pcmq~\psec}}
\newcommand{\ergs}{\mbox{erg~\psec}}
\def\deg{\ensuremath{^\circ}}

\newcommand{\ra}{\mbox{$\alpha_{\rm J2000}$}}
\newcommand{\dec}{\mbox{$\delta_{\rm J2000}$}}
\newcommand{\ecut}{\mbox{$E_{\rm c}$}}
\newcommand{\ext}{\mbox{$\sigma_{\rm ext}$}}
\newcommand{\scut}{\mbox{$s_{\rm c}$}}
\newcommand{\nmsp}{\mbox{$N_{\rm MSP}$}}
\newcommand{\liso}{\mbox{$L_{\gamma}$}}

\newcommand{\edot}{\mbox{$\langle \dot{E} \rangle$}}
\newcommand{\eff}{\mbox{$\langle \eta_{\gamma} \rangle$}}

\newcommand{\Lsol}{\mbox{${\rm L}_{\odot}$}}
\newcommand{\erate}{\mbox{$\Gamma_{\rm e}$}}

\documentclass[longauth]{aa} 

\usepackage{graphicx}
\usepackage{txfonts}
\usepackage{natbib}
\usepackage{lineno}
\bibpunct{(}{)}{;}{a}{}{,} 
\begin{document}

\title{A population of gamma-ray emitting globular clusters seen with the {\em Fermi} Large Area Telescope}

\author{
A.~A.~Abdo$^{1,2}$ \and
M.~Ackermann$^{3}$ \and
M.~Ajello$^{3}$ \and
L.~Baldini$^{4}$ \and
J.~Ballet$^{5}$ \and
G.~Barbiellini$^{6,7}$ \and
D.~Bastieri$^{8,9}$ \and
R.~Bellazzini$^{4}$ \and
R.~D.~Blandford$^{3}$ \and
E.~D.~Bloom$^{3}$
% ...
}
\authorrunning{LAT collaboration}

\institute{
\inst{1}~Space Science Division, Naval Research Laboratory, Washington, DC 20375, USA\\
\inst{2}~National Research Council Research Associate, National Academy of Sciences, Washington, DC 20001, USA\\
\inst{3}~W. W. Hansen Experimental Physics Laboratory, Kavli Institute for Particle Astrophysics and Cosmology, Department of Physics and SLAC National Accelerator Laboratory, Stanford University, Stanford, CA 94305, USA\\
\inst{4}~Istituto Nazionale di Fisica Nucleare, Sezione di Pisa, I-56127 Pisa, Italy\\
\inst{5}~Laboratoire AIM, CEA-IRFU/CNRS/Universit\'e Paris Diderot, Service d'Astrophysique, CEA Saclay, 91191 Gif sur Yvette, France\\
\inst{6}~Istituto Nazionale di Fisica Nucleare, Sezione di Trieste, I-34127 Trieste, Italy\\
\inst{7}~Dipartimento di Fisica, Universit\`a di Trieste, I-34127 Trieste, Italy\\
\inst{8}~Istituto Nazionale di Fisica Nucleare, Sezione di Padova, I-35131 Padova, Italy\\
\inst{9}~Dipartimento di Fisica ``G. Galilei", Universit\`a di Padova, I-35131 Padova, Italy\\
\inst{10}~Istituto Nazionale di Fisica Nucleare, Sezione di Perugia, I-06123 Perugia, Italy\\
\inst{11}~Dipartimento di Fisica, Universit\`a degli Studi di Perugia, I-06123 Perugia, Italy\\
\inst{12}~Centre d'\'Etude Spatiale des Rayonnements, CNRS/UPS, BP 44346, F-31028 Toulouse Cedex 4, France\\
\email{jurgen.knodlseder@cesr.fr}\\
\email{natalie.webb@cesr.fr}\\
\email{benoit.pancrazi@cesr.fr}\\
\inst{13}~Department of Physics, Center for Cosmology and Astro-Particle Physics, The Ohio State University, Columbus, OH 43210, USA\\
\inst{14}~Dipartimento di Fisica ``M. Merlin" dell'Universit\`a e del Politecnico di Bari, I-70126 Bari, Italy\\
\inst{15}~Istituto Nazionale di Fisica Nucleare, Sezione di Bari, 70126 Bari, Italy\\
\inst{16}~Laboratoire Leprince-Ringuet, \'Ecole polytechnique, CNRS/IN2P3, Palaiseau, France\\
\inst{17}~INAF-Istituto di Astrofisica Spaziale e Fisica Cosmica, I-20133 Milano, Italy\\
\inst{18}~George Mason University, Fairfax, VA 22030, USA\\
\inst{19}~Laboratoire de Physique Th\'eorique et Astroparticules, Universit\'e Montpellier 2, CNRS/IN2P3, Montpellier, France\\
\inst{20}~Department of Physics, Stockholm University, AlbaNova, SE-106 91 Stockholm, Sweden\\
\inst{21}~The Oskar Klein Centre for Cosmoparticle Physics, AlbaNova, SE-106 91 Stockholm, Sweden\\
\inst{22}~Royal Swedish Academy of Sciences Research Fellow, funded by a grant from the K. A. Wallenberg Foundation\\
\inst{23}~NASA Goddard Space Flight Center, Greenbelt, MD 20771, USA\\
\inst{24}~Department of Physics and Department of Astronomy, University of Maryland, College Park, MD 20742, USA\\
\inst{25}~CNRS/IN2P3, Centre d'\'Etudes Nucl\'eaires Bordeaux Gradignan, UMR 5797, Gradignan, 33175, France\\
\inst{26}~Universit\'e de Bordeaux, Centre d'\'Etudes Nucl\'eaires Bordeaux Gradignan, UMR 5797, Gradignan, 33175, France\\
\inst{27}~Dipartimento di Fisica, Universit\`a di Udine and Istituto Nazionale di Fisica Nucleare, Sezione di Trieste, Gruppo Collegato di Udine, I-33100 Udine, Italy\\
\inst{28}~Osservatorio Astronomico di Trieste, Istituto Nazionale di Astrofisica, I-34143 Trieste, Italy\\
\inst{29}~Department of Physical Sciences, Hiroshima University, Higashi-Hiroshima, Hiroshima 739-8526, Japan\\
\inst{30}~Agenzia Spaziale Italiana (ASI) Science Data Center, I-00044 Frascati (Roma), Italy\\
\inst{31}~Center for Space Plasma and Aeronomic Research (CSPAR), University of Alabama in Huntsville, Huntsville, AL 35899, USA\\
\inst{32}~Instituci\'o Catalana de Recerca i Estudis Avan\c{c}ats (ICREA), Barcelona, Spain\\
\inst{33}~Research Institute for Science and Engineering, Waseda University, 3-4-1, Okubo, Shinjuku, Tokyo, 169-8555 Japan\\
\inst{34}~Department of Physics, University of Washington, Seattle, WA 98195-1560, USA\\
\inst{35}~Istituto Nazionale di Fisica Nucleare, Sezione di Roma ``Tor Vergata", I-00133 Roma, Italy\\
\inst{36}~Department of Physics and Astronomy, University of Denver, Denver, CO 80208, USA\\
\inst{37}~Hiroshima Astrophysical Science Center, Hiroshima University, Higashi-Hiroshima, Hiroshima 739-8526, Japan\\
\inst{38}~Max-Planck Institut f\"ur extraterrestrische Physik, 85748 Garching, Germany\\
\inst{39}~Institut f\"ur Astro- und Teilchenphysik and Institut f\"ur Theoretische Physik, Leopold-Franzens-Universit\"at Innsbruck, A-6020 Innsbruck, Austria\\
\inst{40}~Santa Cruz Institute for Particle Physics, Department of Physics and Department of Astronomy and Astrophysics, University of California at Santa Cruz, Santa Cruz, CA 95064, USA\\
\inst{41}~NYCB Real-Time Computing Inc., Lattingtown, NY 11560-1025, USA\\
\inst{42}~Department of Chemistry and Physics, Purdue University Calumet, Hammond, IN 46323-2094, USA\\
\inst{43}~Institute of Space and Astronautical Science, JAXA, 3-1-1 Yoshinodai, Sagamihara, Kanagawa 229-8510, Japan\\
\inst{44}~Partially supported by the International Doctorate on Astroparticle Physics (IDAPP) program\\
\inst{45}~Institut de Ciencies de l'Espai (IEEC-CSIC), Campus UAB, 08193 Barcelona, Spain\\
\inst{46}~Consorzio Interuniversitario per la Fisica Spaziale (CIFS), I-10133 Torino, Italy\\
\inst{47}~INTEGRAL Science Data Centre, CH-1290 Versoix, Switzerland\\
\inst{48}~Center for Research and Exploration in Space Science and Technology (CRESST) and NASA Goddard Space Flight Center, Greenbelt, MD 20771, USA\\
\inst{49}~Department of Physics and Center for Space Sciences and Technology, University of Maryland Baltimore County, Baltimore, MD 21250, USA\\
\inst{50}~North-West University, Potchefstroom Campus, Potchefstroom 2520, South Africa\\
\inst{51}~Dipartimento di Fisica, Universit\`a di Roma ``Tor Vergata", I-00133 Roma, Italy\\
\inst{52}~Department of Physics, Royal Institute of Technology (KTH), AlbaNova, SE-106 91 Stockholm, Sweden\\
\inst{53}~School of Pure and Applied Natural Sciences, University of Kalmar, SE-391 82 Kalmar, Sweden\\
\inst{54}~Max-Planck-Institut fŸr Radioastronomie, Auf dem H\"ugel 69, 53121 Bonn, Germany
}

\date{Received March 18, 2010; accepted August 13, 2010}

\abstract
%
{Globular clusters with their large populations of millisecond pulsars (MSPs) are believed to be
potential emitters of high-energy gamma-ray emission.
The observation of this emission provides a powerful tool to assess the millisecond pulsar
population of a cluster, is essential for understanding the importance of binary systems 
for the evolution of globular clusters, and provides complementary insights into magnetospheric
emission processes.}
%
{Our goal is to constrain the millisecond pulsar populations in globular clusters from analysis of 
gamma-ray observations.}
%
{We use 546 days of continuous sky-survey observations obtained with the Large Area 
Telescope aboard the \textit{Fermi} Gamma-ray Space Telescope to study the gamma-ray
emission towards 13 globular clusters.}
%
{Steady point-like high-energy gamma-ray emission has been significantly detected
towards 8 globular clusters.
Five of them (47~Tucanae, Omega~Cen, NGC~6388, Terzan~5, and M~28)
show hard spectral power indices $(0.7 < \Gamma <1.4)$ and clear evidence for an 
exponential cut-off in the range $1.0-2.6$~GeV, which is the characteristic signature
of magnetospheric emission from MSPs.
Three of them (M~62, NGC~6440 and NGC~6652) also show hard spectral indices
$(1.0 < \Gamma < 1.7)$, however the presence of an exponential cut-off can not 
be unambiguously established.
Three of them (Omega Cen, NGC~6388, NGC~6652) have no known radio or
X-ray MSPs yet still exhibit MSP spectral properties.
From the observed gamma-ray luminosities, we estimate the total number of MSPs
that is expected to be present in these globular clusters.
We show that our estimates of the MSP population correlate with the stellar encounter rate and we estimate $2600-4700$ MSPs in Galactic globular clusters, commensurate with previous estimates.}
%
{The observation of high-energy gamma-ray emission from globular clusters thus
provides a reliable independent method to assess their millisecond pulsar populations.}


\maketitle

\section{Introduction}
\label{sec:intro}

With their typical ages of $\sim10^{10}$ years, globular clusters form
the most ancient constituents of our Milky Way Galaxy. These
gravitationally bound concentrations of ten thousand to one million
stars are surprisingly stable against collapse which implies some
source of internal energy that balances gravitation. The potential
energy of binary systems is a plausible source of this internal
energy, tapped by close stellar encounters that harden the orbits of
the systems \citep{hut92}.  Indeed, globular clusters contain
considerably more close binary systems per unit mass than the Galactic
disk which eventually show up as rich populations of X-ray binaries
\citep{clark75}.  This scenario is strengthened by the observation
that the number of low-mass X-ray binary  systems containing neutron
stars is directly correlated with the stellar encounter rate
\citep{gendre03}.  Another consequence of this scenario is the
presence of many millisecond
pulsars\footnote{See
http://www.naic.edu/$\sim$pfreire/GCpsr.html for an updated list.}
\citep[hereafter MSPs; see e.g.][]{camilo05,ransom08},
also known as `recycled' pulsars,  i.e. pulsars that were spun-up to 
millisecond periods by mass-accretion from a binary companion 
\citep{alpar92}.

Observations with the {\em Large Area  Telescope}
(LAT) onboard the {\em Fermi Gamma-ray Space Telescope} have
confirmed MSPs as gamma-ray sources \citep{abdo09a,abdo09b}.  
The spectral energy distribution of millisecond pulsars is characterised
by hard ($1.0\lesssim\Gamma\lesssim2.0$)  power law spectra with
exponential cut-offs in the $1-3$~GeV energy range \citep{abdo09b}.
Recently, \cite{abdo09c} presented the first detection of a globular
cluster (GC) in the gamma-ray domain.  This GC, 47 Tuc, has a spectral energy
distribution best described by a photon index of 1.3$\pm$0.3 with a
cut-off energy of 
$2.5^{+1.6}_{-0.8}$ GeV
\citep{abdo09c} typical of the other MSPs detected to date
\citep{abdo09b}.  Further, 47 Tuc contains at least 23 MSPs, known
from radio and X-ray observations.  The lack of variability over days
to months is consistent with MSP emission.  
In addition, folding the data on known ephemerides from the 47 Tuc pulsars 
reveals no significant detections, thus it appears that the gamma-ray emission is
not due to a single MSP but rather attributable to an entire population of MSPs
in this globular cluster.  
Using the observed, average efficiency of converting spin down energy into the observed 
gamma-ray luminosity, constraints can be placed on the MSP population 
\citep{abdo09c}.

As the number of neutron star X-ray binaries are correlated with
encounter rate and MSPs are the progeny of these systems, it would
follow that the number of MSPs per globular cluster scales in a
similar way.  
It is difficult to test such a correlation using radio and X-ray observations 
as the former are affected by dispersion and scattering by the turbulent ionized 
interstellar medium, in particular for clusters near the Galactic bulge,
while the latter are affected by interstellar absorption rendering the 
detection difficult due to the low count rates observed.

The gamma-ray domain is not affected by interstellar absorption and
there is also the added advantage that the gamma-ray beams may be
wider than the radio/X-ray beams \citep[e.g.][]{abdo10a}, which would
permit more MSPs to be detected in the gamma-ray domain than those at
lower energies, thus making gamma-ray observations ideal for testing
such a correlation.

In this paper we consider gamma-ray sources that are spatially
consistent with GCs and that show the spectral characteristics of
MSPs, i.e. that have hard power law spectra with exponential cut-offs
in the few GeV regime, that are steady, and that are point-like.  We
analyse {\it Fermi} LAT data for 13 globular clusters (see Table
\ref{tab:gc}) and include in our list the 8 globular clusters that
have been formally associated with sources in the first year {\it
  Fermi} LAT catalogue \citep[][hereafter named 1FGL sources]{abdo10b}.
We add two further globular clusters that lie spatially close to 1FGL
sources (Omega Cen and NGC~6624), 
and include also NGC~6441 due to the high stellar collision rate that is 
believed to favour the formation of MSPs \citep{freire08}, 
NGC~6752 due to its relative proximity of 4 kpc \citep{damico02},
and M~15 due to its relatively large population of known MSPs \citep{anderson93}.

\begin{table}[!t]
\footnotesize
\caption{Globular clusters analysed in this work.
\label{tab:gc}
}
\begin{center}
\begin{tabular}{llll}
\hline
\hline
\noalign{\smallskip}
Name & Other name & MSPs & Reason for inclusion \\
\noalign{\smallskip}
\hline
\noalign{\smallskip}
47 Tucanae & NGC~104 & 23 & 1FGL~J0023.9$-$7204 \\
Omega Cen & NGC~5139 & 0 & close to 1FGL~J1328.2$-$4729 \\
M~62 & NGC~6266 & 6 & 1FGL~J1701.1$-$3005 \\
NGC~6388 & ... & 0 & 1FGL~J1735.9$-$4438 \\
Terzan~5 & ... & 33 & 1FGL~J1747.9$-$2448 \\
NGC~6440 & ... & 6 & 1FGL~J1748.7$-$2020 \\
NGC~6441 & ... & 4 & high collision rate \\
NGC~6541 & ... & 0 & 1FGL~J1807.6$-$4341 \\
NGC~6624 & ... & 4 & close to 1FGL J1823.4$-$3009 \\
M~28 & NGC~6626 & 12 & 1FGL~J1824.5$-$2449 \\
NGC~6652 & ... & 0 & 1FGL~J1835.3$-$3255 \\
NGC~6752 & ... & 5 & 5 MSPs, nearby \\
M~15 & NGC~7078 & 8 & 8 MSPs \\
\noalign{\smallskip}
\hline
\end{tabular}
\end{center}
{\bf Notes.}
The known number of MSPs (column 3) has been taken from
http://www.naic.edu/$\sim$pfreire/GCpsr.html.
\end{table}

\section{Observations}
\label{sec:observation}

\subsection{Data preparation}

The data used in this work have been acquired by the LAT telescope aboard the
{\em Fermi} Gamma-ray Space Telescope during continuous regular sky survey covering
the period August 8th 2008 -- February 12th 2010 (546 days).
Events satisfying the standard low-background event
selection \citep[`Diffuse' events;][]{atwood09} and coming from zenith angles
$<105\deg$ (to greatly reduce the contribution by Earth albedo gamma rays)
were used.
Furthermore, we selected only events where the satellite rocking angle was 
inferior to $40\deg$ for the first 12 months, and inferior to $52\deg$ for the remaining 
period.
We further restricted the analysis to photon energies above 200~MeV; below this energy,
the effective area in the `Diffuse class' is relatively small and strongly dependent on 
energy.
All analysis was performed using the LAT Science Tools package, which is available 
from the Fermi Science Support Center, using P6\_V3 post-launch instrument response 
functions (IRFs).
These take into account pile-up and accidental coincidence effects in the
detector subsystems that were not considered in the definition of the pre-launch
IRFs.

\subsection{Analysis method}
\label{sec:method}

For the analysis of each globular cluster we selected events within squared 
regions-of-interest (ROIs) that have been aligned in Galactic coordinates.
Events have been binned into 25 logarithmically-spaced energy bins covering the
range 200 MeV - 50 GeV and into spatial pixels of $0.1\deg \times 0.1\deg$ in size
using a Cartesian projection.
The size and location of the ROIs have been chosen to avoid nearby strong sources
and the bright diffuse emission from the Galactic plane while maintaining the largest
possible coverage of the point-spread function for the globular clusters of interest.
The ROI definitions are summarised in Table~\ref{tab:roi}.

\begin{table}[!t]
\footnotesize
\caption{Definition of ROIs used for analysis.
\label{tab:roi}
}
\begin{center}
\begin{tabular}{lrrr}
\hline
\hline
\noalign{\smallskip}
Name & \multicolumn{1}{c}{\ra} & \multicolumn{1}{c}{\dec} & \multicolumn{1}{c}{Size} \\
\noalign{\smallskip}
\hline
\noalign{\smallskip}
47 Tucanae & $00^{\rm h}24^{\rm m}01.7^{\rm s}$ & $-72\deg04'42.9''$ & $10\deg \times 10\deg$ \\
Omega Cen & $13^{\rm h}26^{\rm m}45.0^{\rm s}$ & $-47\deg28'37.0''$ & $10\deg \times 10\deg$ \\
M~62            & $17^{\rm h}01^{\rm m}10.3^{\rm s}$ & $-30\deg05'00.6''$ & $6\deg \times 6\deg$ \\
NGC~6388  & $17^{\rm h}36^{\rm m}17.0^{\rm s}$ & $-44\deg44'05.6''$ & $6\deg \times 6\deg$ \\
Terzan~5     & $17^{\rm h}48^{\rm m}04.9^{\rm s}$ & $-24\deg46'48.0''$ & $6\deg \times 6\deg$ \\
NGC~6440  & $17^{\rm h}48^{\rm m}52.0^{\rm s}$ & $-20\deg21'34.0''$ & $6\deg \times 6\deg$ \\
NGC~6441  & $17^{\rm h}50^{\rm m}12.0^{\rm s}$ & $-37\deg03'04.0''$ & $6\deg \times 6\deg$ \\
NGC~6541  & $18^{\rm h}08^{\rm m}02.2^{\rm s}$ & $-43\deg42'19.0''$ & $6\deg \times 6\deg$ \\
NGC~6624  & $18^{\rm h}23^{\rm m}40.0^{\rm s}$ & $-30\deg21'38.0''$ & $6\deg \times 6\deg$ \\
M~28            & $18^{\rm h}23^{\rm m}46.0^{\rm s}$ & $-24\deg55'19.0''$ & $6\deg \times 6\deg$ \\
NGC~6652  & $18^{\rm h}35^{\rm m}45.0^{\rm s}$ & $-32\deg59'25.0''$ & $6\deg \times 6\deg$ \\
NGC~6752  & $19^{\rm h}10^{\rm m}51.0^{\rm s}$ & $-59\deg58'54.0''$ & $10\deg \times 10\deg$ \\
M~15            & $21^{\rm h}29^{\rm m}58.0^{\rm s}$ & $+12\deg10'00.4''$ & $10\deg \times 10\deg$ \\
\noalign{\smallskip}
\hline
\end{tabular}
\end{center}
{\bf Notes.}
Columns are
(1) source name,
(2) Right Ascension (J2000) of ROI centre,
(3) Declination (J2000) of ROI centre, and
(4) squared size of ROI.
\end{table}

The observed events in each of the ROIs have been modelled using components for
the gamma-ray and instrumental backgrounds and for known sources in the field.
The backgrounds are a combination of extragalactic and Galactic diffuse 
emissions and some residual instrumental background.
The Galactic component has been modelled using the LAT standard diffuse background 
model {\tt gll\_iem\_v02.fit} for which we kept the normalisation as a free parameter.
The extragalactic and residual instrumental backgrounds have been combined into a 
single component which has been taken as being isotropic.
The spectrum of this component has been modelled using the tabulated model
{\tt isotropic\_iem\_v02.txt} and the normalisation 
has been kept as a free parameter.\footnote{
The models can be downloaded from
http://fermi.gsfc.nasa.gov/ssc/data/access/lat/BackgroundModels.html.
}
In addition to the diffuse components, we included all point sources 
(except for the source that corresponds to the globular cluster of interest) 
from the first-year 
catalogue \citep{abdo10b} that were detected with TS~$\ge25$ for a given ROI in the
background model.
As usual, the {\em Test Statistic} TS is defined as twice the difference 
between the log-likelihood with ($\mathcal{L}_1$) and without ($\mathcal{L}_0$)
the source, i.e. ${\rm TS} = 2(\mathcal{L}_1 - \mathcal{L}_0)$
(TS~$=25$ corresponds to a detection significance of $\sim4\sigma$; \citet{abdo10b}).
In general, the spectral distributions of the point sources have been fitted using
power law models for which we kept the total flux and the spectral index as free
parameters.
Only if a point source has been identified as a gamma-ray pulsar, we fitted its spectral
distribution using an exponentially cut-off power law for which we kept the normalisation,
the spectral index and the cut-off energy as free parameters.

As a first analysis step, we localised the gamma-ray emission by fitting a point source on 
top of the background model for a grid of test positions 
using the binned maximum likelihood fitting procedure {\tt gtlike}.
For each ROI, the grid has been centred on the nominal position of the globular 
cluster\footnote{Taken from SIMBAD.} and was comprised of $20 \times 20$ positions 
with a spacing of $3'$.
We modelled the spectrum of the point source using a simple power law for which we
kept the integrated flux and the spectral index as free parameters.
The TS values obtained for the point source at the grid positions were then interpolated
using a minimum curvature surface method to approximate the TS surface near the 
globular cluster.
This surface was then used to localise TS$_{\rm max}$, the position $(\ra, \dec)$ at which 
TS is maximised.
The $95\%$ confidence region for the true source location was derived from the
surface contour that fulfils ${\rm TS}_{\rm max}-6.0$.
For simplicity, we approximated this contour by a circle with radius $r_{95}$.

As next step, we determined the spatial extent of the gamma-ray emission by replacing
the point source in the grid search by 2D Gaussian-shaped intensity profiles and by
repeating the search for various Gaussian extents \ext.
Typically, we performed the grid search for $\sigma_{\rm ext}=0.1\deg$,  
$0.2\deg$, and $0.3\deg$, but eventually we also performed runs for 
$\sigma_{\rm ext}=0.05\deg$ and $0.4\deg$ to better constrain the source
extent.
The dependence of TS$_{\rm max}$ on \ext\ was then determined by adjusting
a parabola to the pairs $({\rm TS}_{\rm max}, \ext)$ of values, which in all cases
gave a satisfactory fit near the maximum.
Using this parabola we then determined the maximum TS$_{\rm max}$ and the
corresponding \ext\ as the maximum likelihood estimate of the source extent.
All the LAT sources were consistent with being point 
sources, and we determined the $2\sigma$ upper extension limit from the parabola
by searching the \ext\ for which the TS decreased by 4 from its maximum value.
If the optimum \ext\ was formally $>0$, we added this (generally small)
offset to the upper limit.

Once we established that the gamma-ray source is spatially consistent with the globular
cluster and that the source is not extended, we determined the spectral characteristics
of the source by fitting a point source at the nominal position of the globular cluster
on top of the background model to the data.
We again used the binned maximum likelihood fitting procedure {\tt gtlike} for this
purpose.
For the spectral model of the source we used an exponentially cut-off power law
$N(E) = N_0 E^{-\Gamma} e^{-E/E_{\rm c}}$,
with normalisation $N_0$, spectral index $\Gamma$, and cut-off energy \ecut.
We estimated the significance \scut\ of the cut-off by repeating the fit using
a simple power law model, and by using
$s_{\rm c}=\sqrt{{\rm TS}_{\rm c} - {\rm TS}_{\rm p}}$, where
${\rm TS}_{\rm c}$ and ${\rm TS}_{\rm p}$ are the TS values obtained using the
exponentially cut-off power law and the simple power law, respectively.

We determined 68\% confidence intervals for all spectral parameters by successively
varying each parameter while optimising the others until
${\rm TS}={\rm TS}_{\rm max}-1.0$
is fulfilled.
The resulting statistical parameter uncertainties are to first order consistent with the 
values that we obtained using {\tt gtlike}, which are estimated from the covariance 
matrix of the parameters at maximum likelihood.
While the latter uncertainties are by definition symmetric about the maximum likelihood
parameter values, our method allows for the determination of asymmetric confidence 
intervals, which are particularily relevant for describing the statistical uncertainty in 
the measurements of \ecut.

In addition to the statistical uncertainties we also estimated systematic uncertainties for
all spectral parameters that arise from uncertainties in the precise knowledge of the 
effective area of the LAT telescope and of its energy dependence.
We do this by repeating the {\tt gtlike} fits using modified instrument response functions 
that bracket the uncertainties in the effective area.
For each parameter we quote as systematic uncertainty the maximum parameter offset
that is obtained with respect to the analysis done with the P6\_V3 IRF.

We also searched for possible flux variability of the sources associated with the globular
cluster by fitting our source model to the data on a monthly basis.
For this analysis we fixed the power law spectral indices of all sources and the
normalisation of the isotropic diffuse component to the values we obtained for the
full dataset.
The normalisation of the Galactic diffuse component was kept as a free parameter
due to the fact that this component was found to vary significantly, in particular for 
ROIs that are close to the Galactic plane.
From this analysis we obtained lightcurves for all globular clusters that we fitted to 
a constant flux model using a $\chi^2$ minimisation procedure.
Since we fit lightcurves for 17 months of data, we have 16 degrees-of-freedom in this
minimisation, and the source can be considered as not being constant at the
$68\%$ or $95\%$ confidence levels if $\chi^2_{\rm month} >18.1$ or $>26.3$, 
respectively.

Finally, we derived the spectral energy distributions (SEDs) of the globular clusters 
by dividing the energy range from 200 MeV to 20 GeV into 8 logarithmically-spaced 
energy bins, and by fitting the fluxes or normalisations of all model components 
independently for each bin.
For all sources, we assumed simple power law models with a fixed spectral
index of $\Gamma=2.1$ within each energy bin.
Since the energy bins chosen are relatively narrow, the precise value of the slope has 
little impact on the resulting spectra.
The normalisations of the diffuse components (Galactic diffuse and isotropic backgrounds) 
are fitted independently for each energy bin, which makes this analysis less sensitive
to systematic uncertainties in the spectral energy distribution of these model components.

We also note that we searched for pulsations from MSPs in globular clusters and, to date, 
no significant pulsations have been reported from any of the pulsars that has been
searched.
However, we only have accurate ephemerides (derived from times of arrival measured
with the Parkes, Nancay, and Jodrell Bank radio telescopes) for 15 globular cluster 
MSPs, limiting our search to only $\sim10\%$ of the known globular cluster pulsar
population. 
This illustrates the pressing need for more contemporaneous timing solutions in order 
to reveal the gamma-ray pulsations from globular cluster MSPs.

\subsection{Results}

\begin{figure*}[!ht]
\centering
\includegraphics[width=6cm]{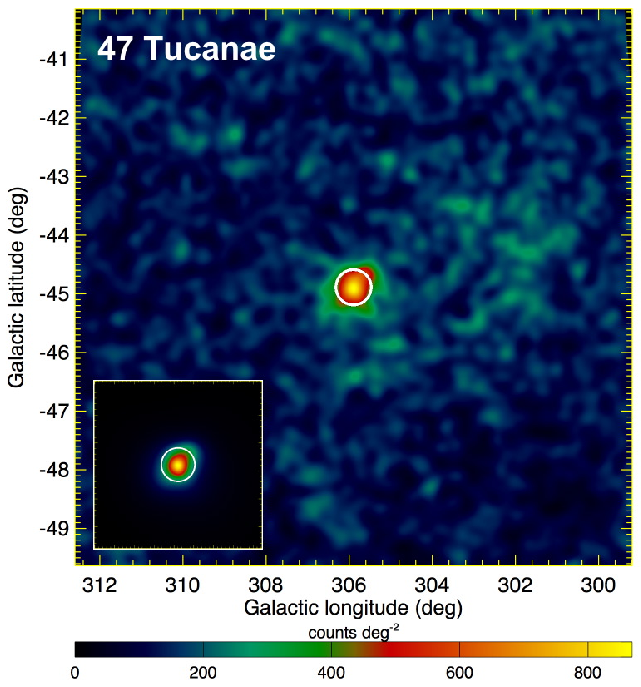}
\includegraphics[width=6cm]{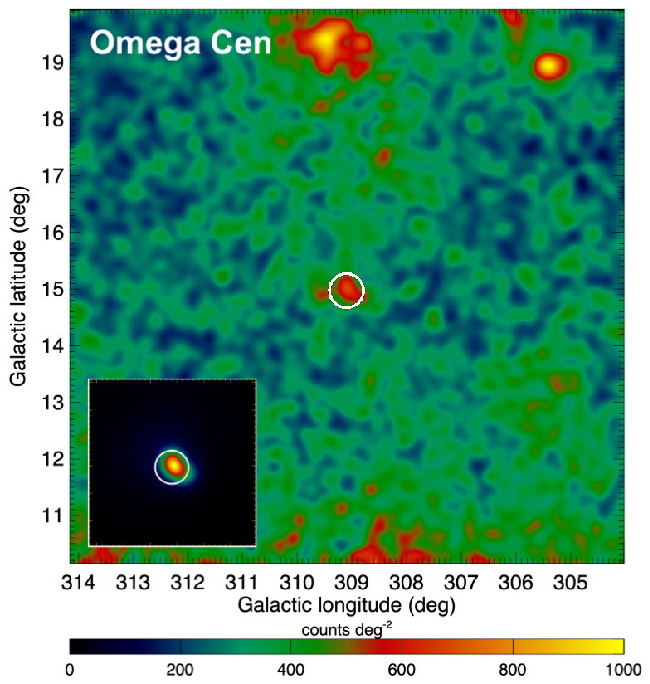}
\includegraphics[width=6cm]{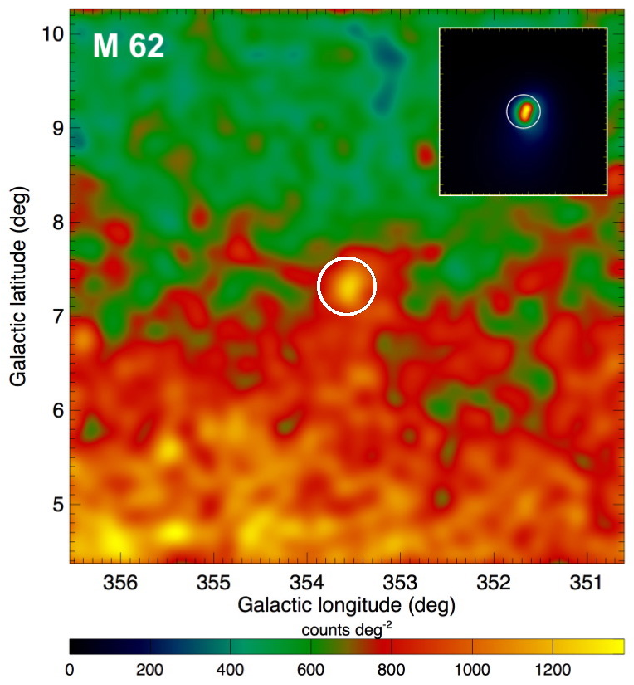}
\includegraphics[width=6cm]{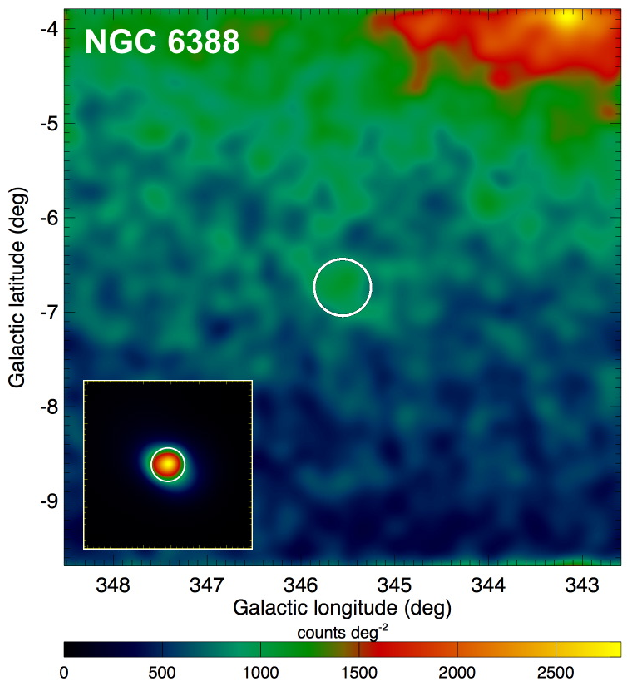}
\includegraphics[width=6cm]{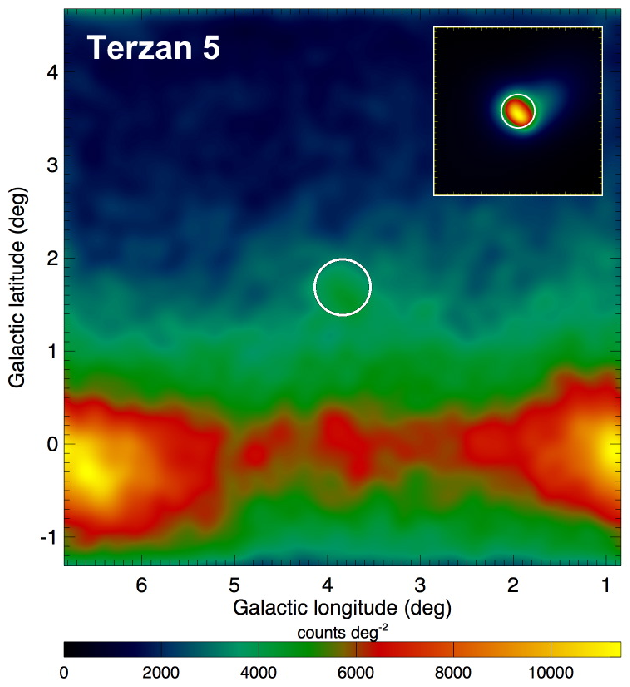}
\includegraphics[width=6cm]{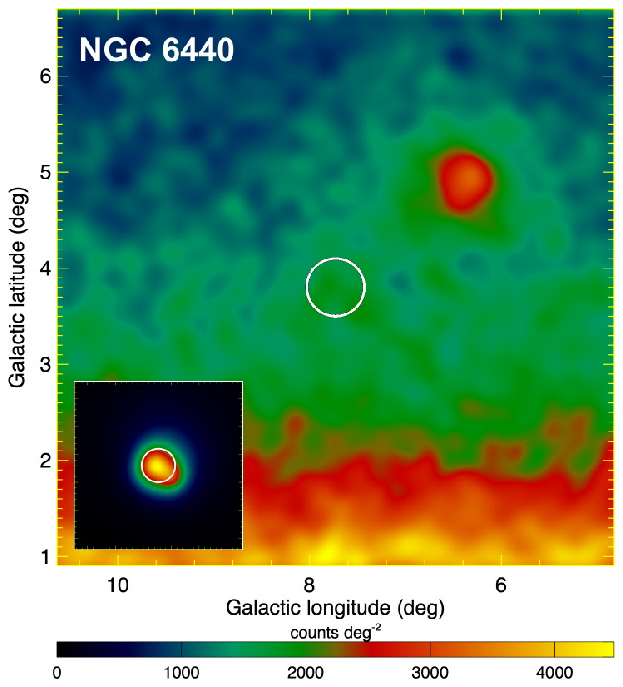}
\includegraphics[width=6cm]{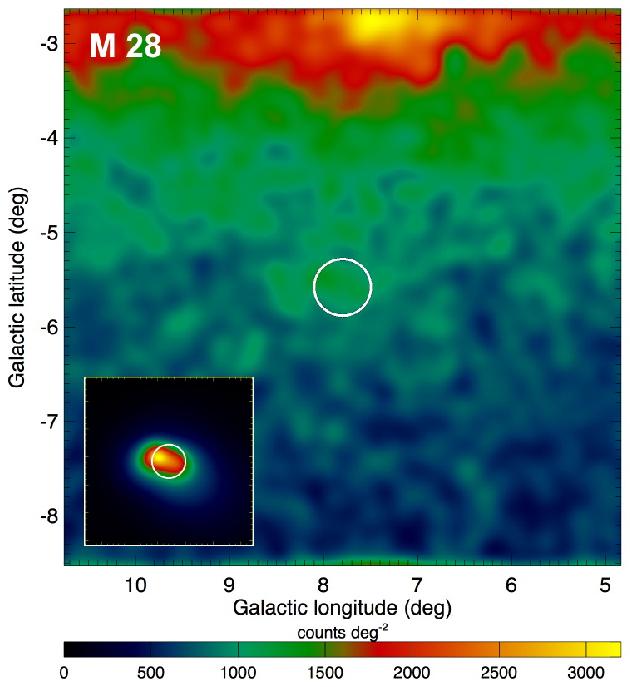}
\includegraphics[width=6cm]{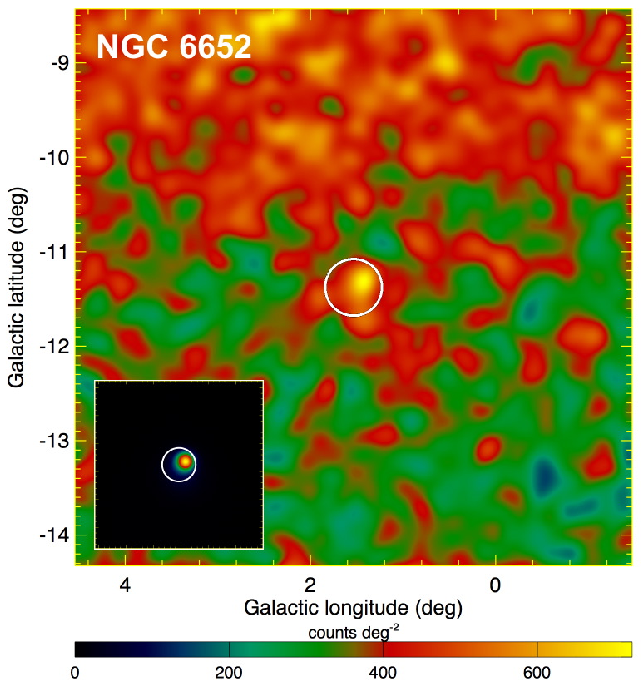}
\hfill
\hfill
\caption{
Gaussian kernel smoothed ($\sigma=0.1\deg$) counts maps of the ROIs for the 8 significant
globular cluster candidates.
The insets show adaptively smoothed background subtracted counts maps for a
$3\deg \times 3\deg$ large region around the globular cluster.
Circles indicate the nominal locations of the globular clusters.
\label{fig:images}
}
\end{figure*}

For 11 of the 13 ROIs analysed, a gamma-ray source has been found in the 
{\it Fermi} LAT data that is spatially consistent with a globular cluster.
For two ROIs (NGC~6441 and NGC~6624) we were unable to find a convincing gamma-ray 
counterpart; however, gamma-ray sources were detected close by to both globular 
clusters.
For NGC~6441 we found evidence for a faint gamma-ray source (TS~$=29$) that we 
localised at 
$\ra= 17^{\rm h}50.5^{\rm m}$ and $\dec=-36\deg52'$ with $r_{95}=7.5'$, yet with an offset
of $11.7'$, the globular cluster is located just outside the $99\%$ error contour.
NGC~6624 is close to 1FGL J1823.4$-$3009 \citep{abdo10b} which we detected with
TS~$=43$, and which we localised at
$\ra= 18^{\rm h}23.5^{\rm m}$ and $\dec=-30\deg13'$ with $r_{95}=6.3'$.
With an offset of $8.6'$, NGC~6624 is also located just outside the $99\%$ error contour
of the gamma-ray source.
Formally, we can not establish an association of NGC~6441 and NGC~6624 with
the gamma-ray sources, and hence we excluded both objects from further analysis.
Both gamma-ray sources, however, are rather faint, and more data are needed before
an association of these sources with globular clusters definitely can be excluded.

Three of the 11 gamma-ray sources that are spatially consistent with a globular cluster 
have TS~$<25$ (NGC~6541, NGC~6752, and M~15), hence we cannot claim a significant
detection for these cases.
NGC~6541 has been associated in the first year catalogue with 1FGL~J1807.6$-$4341
\citep{abdo10b}, yet with ${\rm TS}=25.3$ this source was just barely above the threshold
of ${\rm TS}=25$ used for compilation of the catalogue.
In our dedicated analysis we find a lower detection significance of ${\rm TS}=12.0$
for the source, which can be explained by differences in the analysis method 
and data selection with respect to the catalogue analysis.
In particular, the catalogue analysis determined TS using an unbinned maximum 
likelihood analysis which has recently been recognized to provide larger TS values 
in several low-latitude fields than the binned analysis

This leaves us with 8 significant globular cluster candidates for which we show smoothed 
raw and residual counts maps in Fig.~\ref{fig:images}.
For each of these clusters we determined fluxes and spectral parameters using 
exponentially cut-off power laws.
For the 3 non-significant globular clusters we derived formal $2\sigma$ upper flux limits 
by fitting a point source at the nominal cluster position to the data for which we increased 
the flux, starting from its maximum likelihood value, until the TS decreased by $4$ with 
respect to the maximum.
As a spectral model we used an exponentially cut-off power law for which we fixed the
spectral index and the cut-off energy to the weighted average values of $\Gamma=1.4$ and
$\ecut=1.6$~GeV that were obtained for the 8 significantly detected globular cluster
candidates.
We added to these statistical upper limits the systematic uncertainty that has been obtained
using the bracketing IRFs (cf.~Section \ref{sec:method}) and that accounts for
uncertainties in the effective area of the LAT.
The fluxes and spectral characteristics as well as the maximum likelihood locations of the
gamma-ray sources are summarised in Table \ref{tab:results}.
Quoted uncertainties are statistical and systematic.

\begin{table*}[!th]
\footnotesize
\caption{Gamma-ray characteristics of globular clusters.
\label{tab:results}
}
\begin{center}
\begin{tabular}{lrrrrrrccccc}
\hline
\hline
\noalign{\smallskip}
Name & \multicolumn{1}{c}{\ra} & \multicolumn{1}{c}{\dec} & \multicolumn{1}{c}{$r_{95}$} &
\multicolumn{1}{c}{$\sigma_{\rm ext}$} & 
 \multicolumn{1}{c}{TS} & \multicolumn{1}{c}{$\chi^2_{\rm month}$} & 
 Photon flux & Energy flux & $\Gamma$ & $E_{\rm c}$ &
 $s_{\rm c}$ \\
\noalign{\smallskip}
\hline
\noalign{\smallskip}
47 Tucanae &
  $00^{\rm h}23.8^{\rm m}$ &  
  $-72\deg04'$ & 
  $3.3'$ &
  $<4.8'$ &
  $603.3$ &
  $9.6$ &
  $2.9^{+0.6+0.4}_{-0.5-0.3}$ & 
  $2.5^{+0.2+0.2}_{-0.2-0.2}$ & 
  $1.4^{+0.2+0.2}_{-0.2-0.2}$ & 
  $2.2^{+0.8+0.3}_{-0.5-0.2}$ &
  $5.6$ \\
\noalign{\smallskip}
Omega Cen &
  $13^{\rm h}26.5^{\rm m}$ & 
  $-47\deg29'$ & 
  $7.5'$ &
  $<8.4'$ &
  $50.0$ &
  $14.6$ &
  $0.9^{+0.5+0.3}_{-0.4-0.2}$ & 
  $1.0^{+0.2+0.1}_{-0.2-0.1}$ & 
  $0.7^{+0.7+0.4}_{-0.6-0.4}$ & 
  $1.2^{+0.7+0.2}_{-0.4-0.2}$ &
  $4.0$ \\
\noalign{\smallskip}
M~62 &
  $17^{\rm h}01.1^{\rm m}$ & 
  $-30\deg08'$ & 
  $4.4'$ &
  $<7.2'$ &
  $107.9$ &
  $16.0$ &
  $2.7^{+1.0+1.9}_{-0.9-0.8}$ & 
  $2.1^{+0.3+0.5}_{-0.3-0.1}$ & 
  $1.7^{+0.3+0.4}_{-0.3-0.5}$ & 
  $4.4^{+3.8+17.7}_{-1.8-1.8}$ &
  $2.5$ \\
\noalign{\smallskip}
NGC~6388 &
  $17^{\rm h}35.9^{\rm m}$ & 
  $-44\deg41'$ & 
  $5.7'$ &
  $<9.0'$ &
  $86.6$ &
  $13.8$ &
  $1.6^{+1.0+2.0}_{-0.6-0.6}$ & 
  $1.6^{+0.3+0.6}_{-0.3-0.2}$ & 
  $1.1^{+0.7+0.8}_{-0.5-0.8}$ & 
  $1.8^{+1.2+1.8}_{-0.7-0.6}$ &
  $3.3$ \\
\noalign{\smallskip}
Terzan~5 &
  $17^{\rm h}47.9^{\rm m}$ & 
  $-24\deg48'$ & 
  $2.9'$ &
  $<9.0'$ &
  $341.3$ &
  $25.5$ &
  $7.6^{+1.7+3.4}_{-1.5-2.2}$ & 
  $7.1^{+0.6+1.0}_{-0.5-0.5}$ & 
  $1.4^{+0.2+0.4}_{-0.2-0.3}$ & 
  $2.6^{+0.7+1.2}_{-0.5-0.7}$ &
  $7.1$ \\
\noalign{\smallskip}
NGC~6440 & 
  $17^{\rm h}48.8^{\rm m}$ & 
  $-20\deg21'$ & 
  $5.2'$ &
  $<8.4'$ &
  $65.7$ &
  $5.9$ &
  $2.9^{+2.7+4.4}_{-1.3-1.1}$ & 
  $2.2^{+0.9+1.2}_{-0.5-0.2}$ & 
  $1.6^{+0.5+0.6}_{-0.5-0.8}$ & 
  $3.1^{+3.3+\infty}_{-1.4-1.1}$ &
  $1.4$ \\
\noalign{\smallskip}
M~28 &
  $18^{\rm h}24.4^{\rm m}$ & 
  $-24\deg51'$ & 
  $8.0'$ &
  $<15.6'$ &
  $77.9$ &
  $20.6$ & 
  $2.6^{+1.3+2.2}_{-1.0-0.9}$ & 
  $2.0^{+0.4+0.6}_{-0.3-0.3}$ & 
  $1.1^{+0.7+0.6}_{-0.5-0.7}$ & 
  $1.0^{+0.6+0.4}_{-0.3-0.2}$ &
  $4.3$ \\
\noalign{\smallskip}
NGC~6652 &
  $18^{\rm h}35.7^{\rm m}$ & 
  $-33\deg01'$ & 
  $7.5'$ &
  $<9.6'$ &
  $54.8$ &
  $9.8$ & 
  $0.7^{+0.5+0.2}_{-0.3-0.1}$ & 
  $0.8^{+0.2+0.1}_{-0.1-0.1}$ & 
  $1.0^{+0.6+0.3}_{-0.5-0.3}$ & 
  $1.8^{+1.2+0.4}_{-0.6-0.3}$ &
  $3.2$ \\
\noalign{\smallskip}
\hline
\noalign{\smallskip}
NGC~6541 &
  $18^{\rm h}07.9^{\rm m}$ & 
  $-43\deg41'$ & 
  $20.1'$ &
  ... &
  $12.0$ &
  ... &
  $<1.1$ & 
  $<0.8$ & 
  $(1.4)$ & 
  $(1.6)$ &
  ... \\
NGC~6752 &
  $19^{\rm h}10.3^{\rm m}$ & 
  $-59\deg56'$ & 
  $6.3'$ & 
  ... &
  $13.7$ &
  ... &
  $<0.7$ & 
  $<0.5$ & 
  $(1.4)$ & 
  $(1.6)$ &
  ... \\
M~15 &
  $21^{\rm h}29.4^{\rm m}$ & 
  $+12\deg06'$ & 
  $6.9'^{\ast}$ & 
  ... &
  $5.4$ &
  ... &
  $<0.6$ & 
  $<0.5$ & 
  $(1.4)$ & 
  $(1.6)$ &
  ... \\
\noalign{\smallskip}
\hline
\end{tabular}
\end{center}
{\bf Notes.}
Columns are
(1) source name,
(2) fitted Right Ascension (J2000) of LAT source,
(3) fitted Declination (J2000) of LAT source,
(4) 95\% confidence level error radius 
  ($^{\ast}$for M~15 the 68\% confidence level error radius is quoted because the small
  significance of the detection does not allow to define a meaningful 95\% confidence
  level error radius),
(5) $2\sigma$ upper limit on source extent, defined as $\sigma$ of a 2D Gaussian intensity
  profile,
(6) value of Test Statistic,
(7) $\chi^2_{\rm month}$ as a measure of source variability,
(8) integrated $>100$~MeV photon flux in units of $10^{-8}$ \funit,
(9) integrated $>100$~MeV energy flux in units of $10^{-11}$ \eunit,
(10) spectral index $\Gamma$ of exponential cut-off power law, assuming 
$N(E) \propto E^{-\Gamma} e^{-E/E_{\rm c}}$,
(11) cut-off energy $E_{\rm c}$ in GeV, and
(12) significance of exponential cut-off in units of Gaussian $\sigma$.
\end{table*}

\begin{figure*}[!ht]
\centering
\includegraphics[width=9cm]{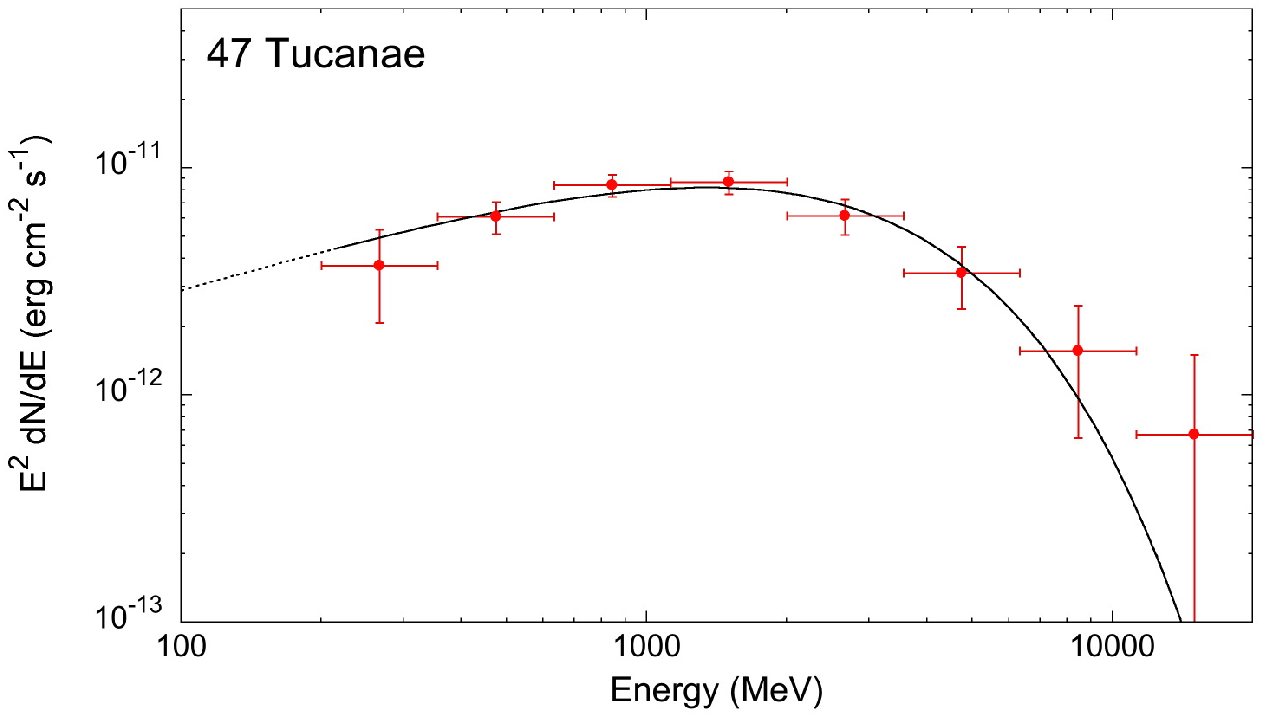}
\includegraphics[width=9cm]{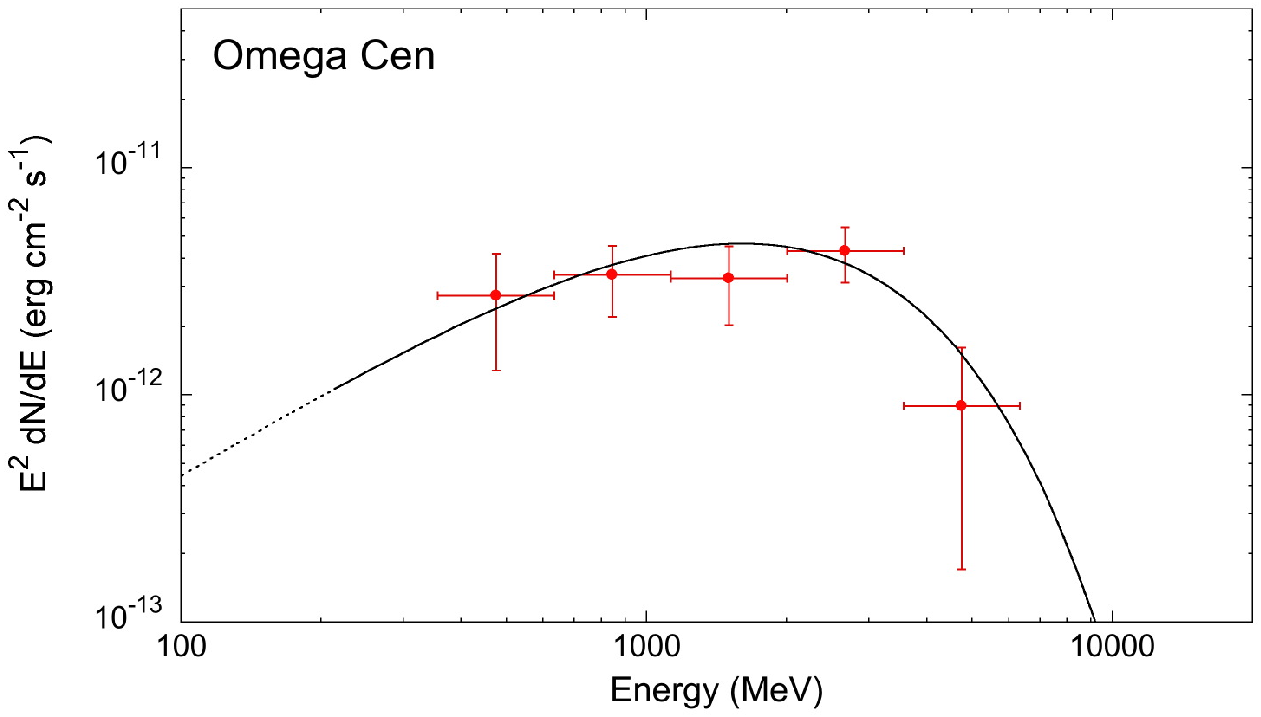}
\includegraphics[width=9cm]{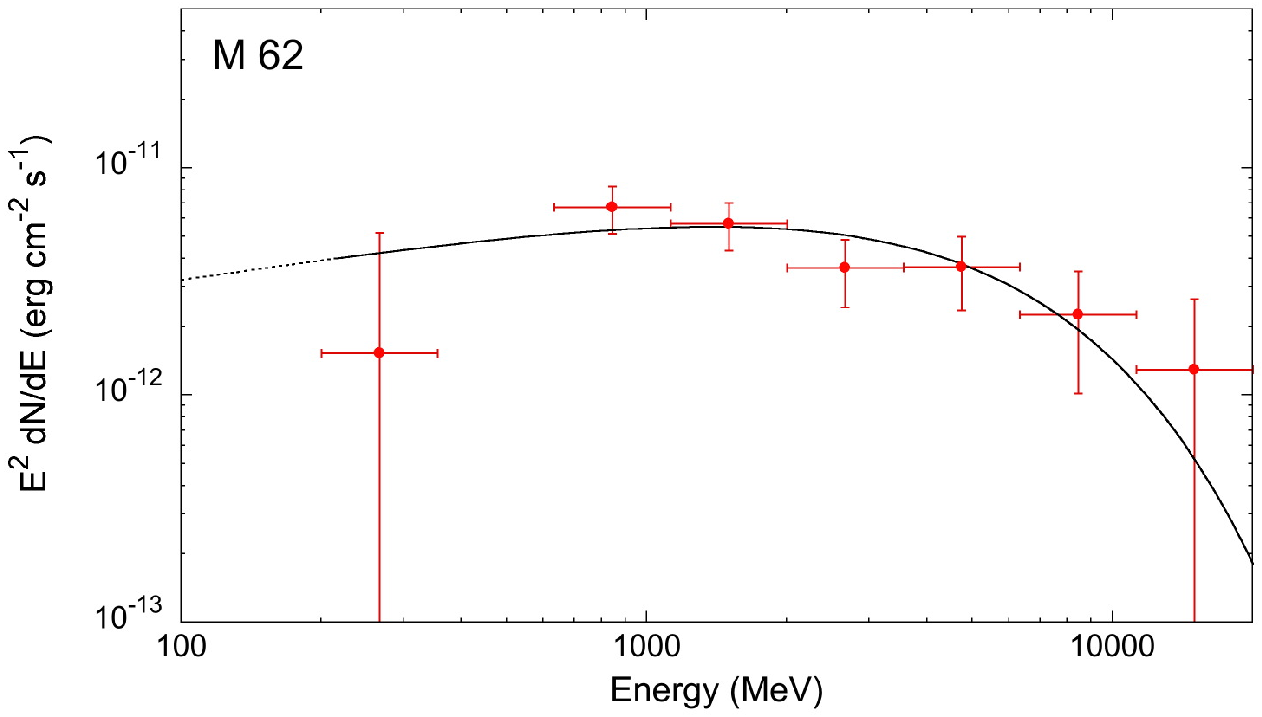}
\includegraphics[width=9cm]{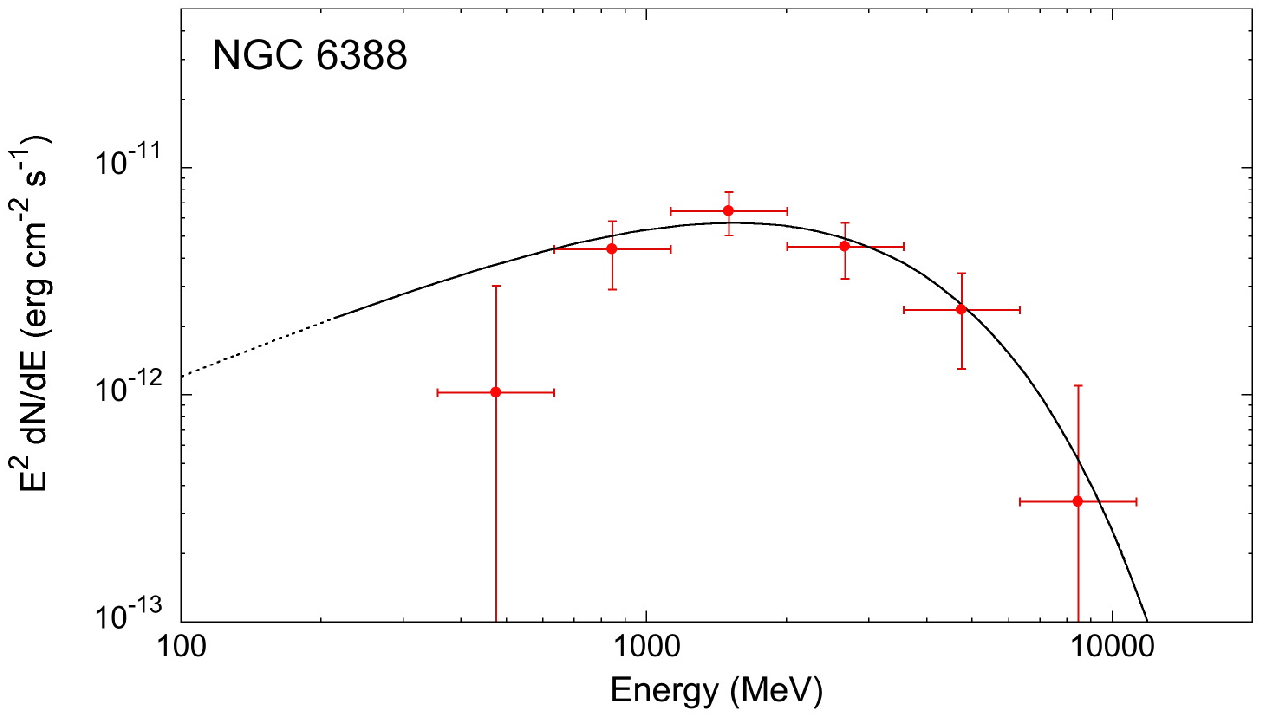}
\includegraphics[width=9cm]{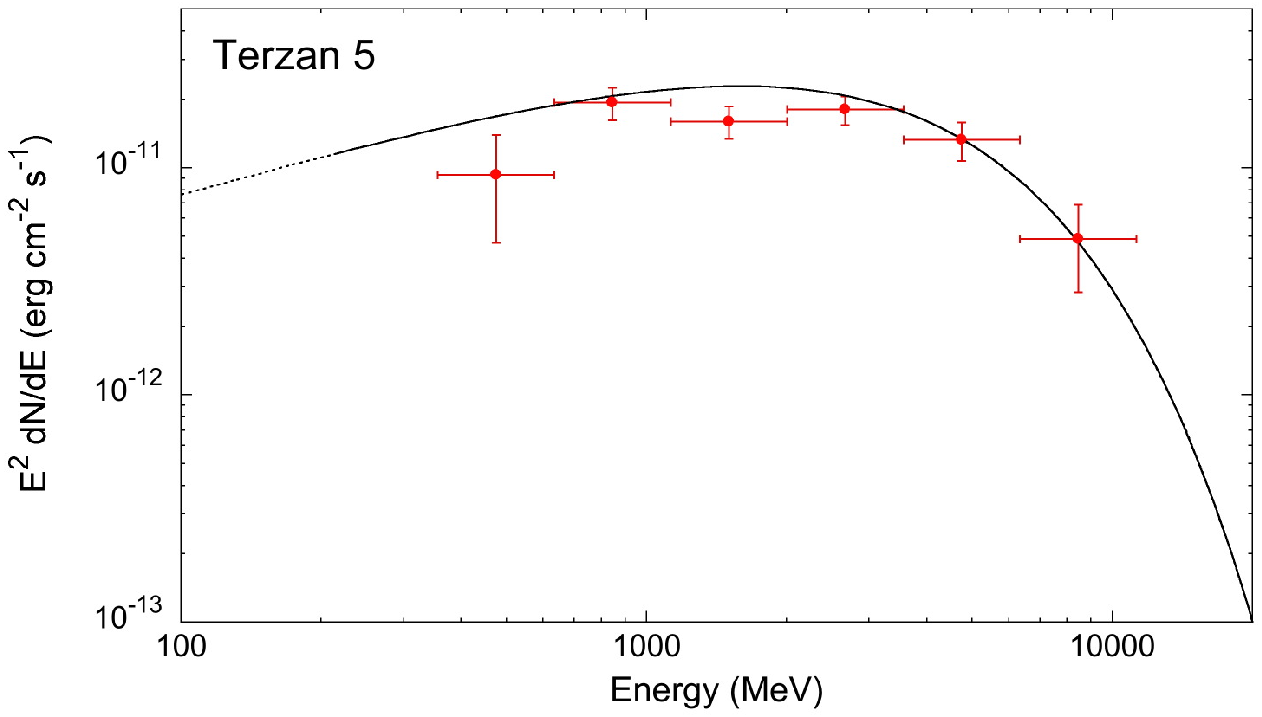}
\includegraphics[width=9cm]{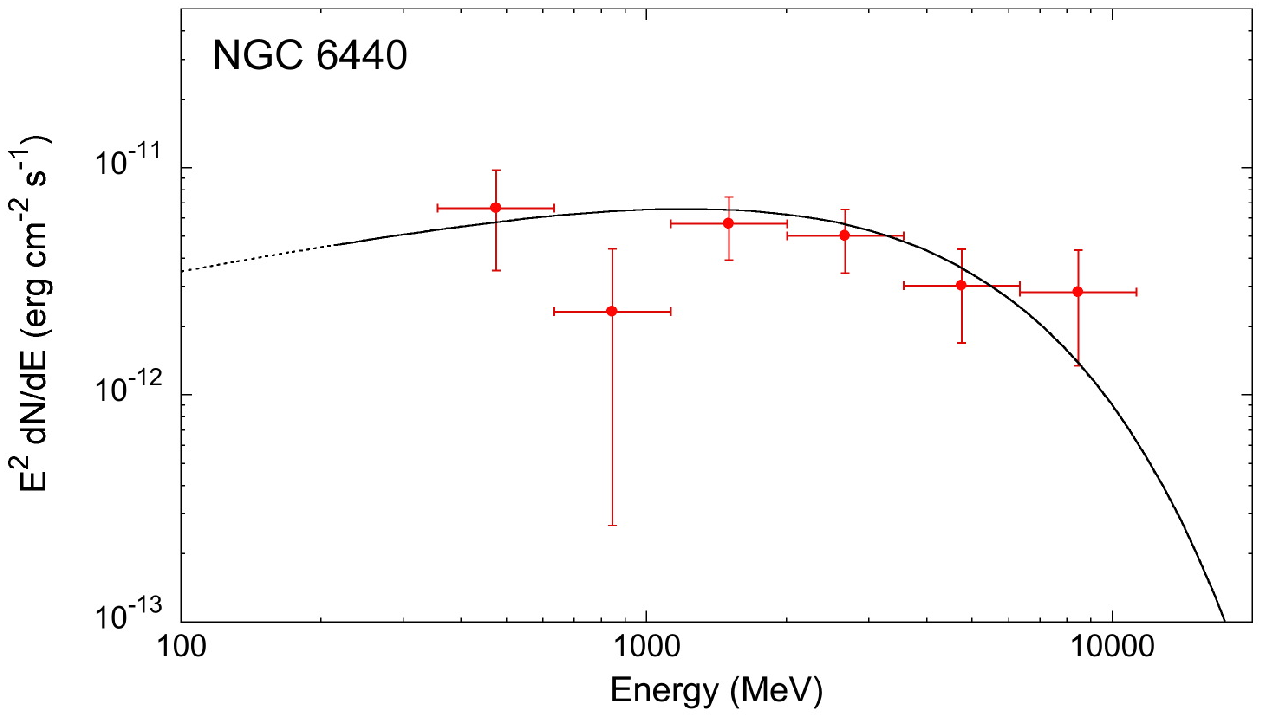}
\includegraphics[width=9cm]{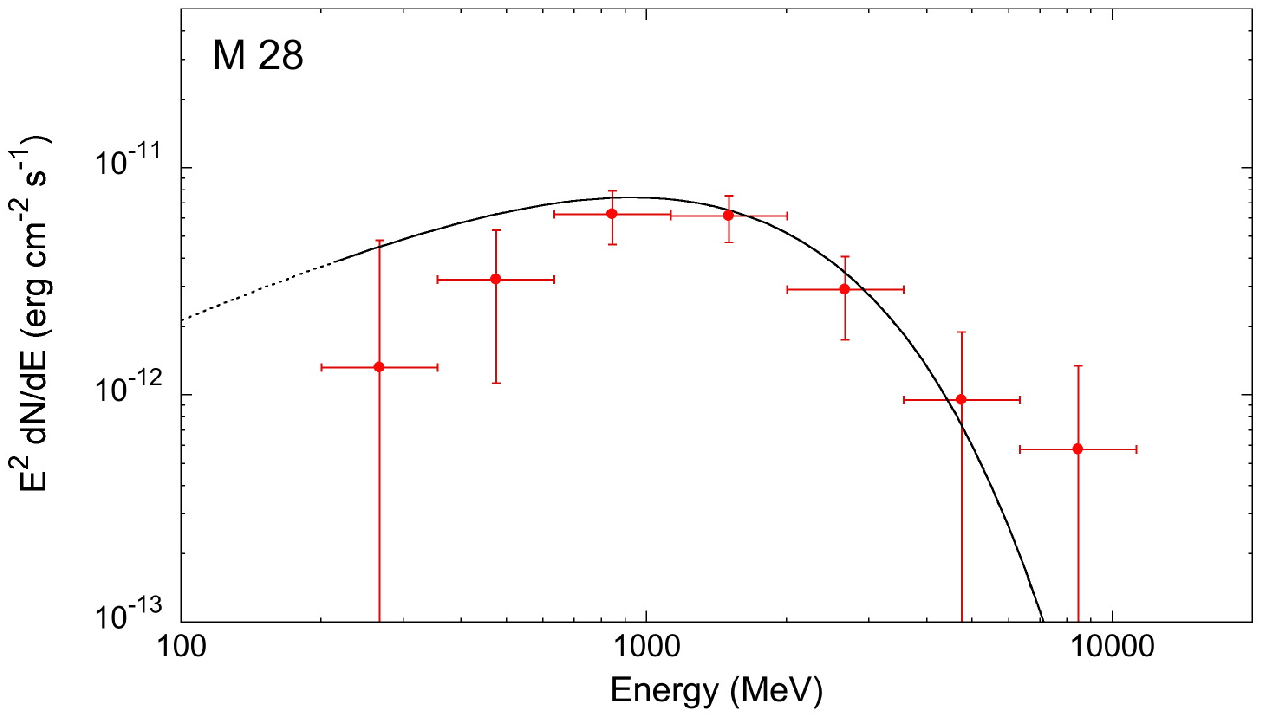}
\includegraphics[width=9cm]{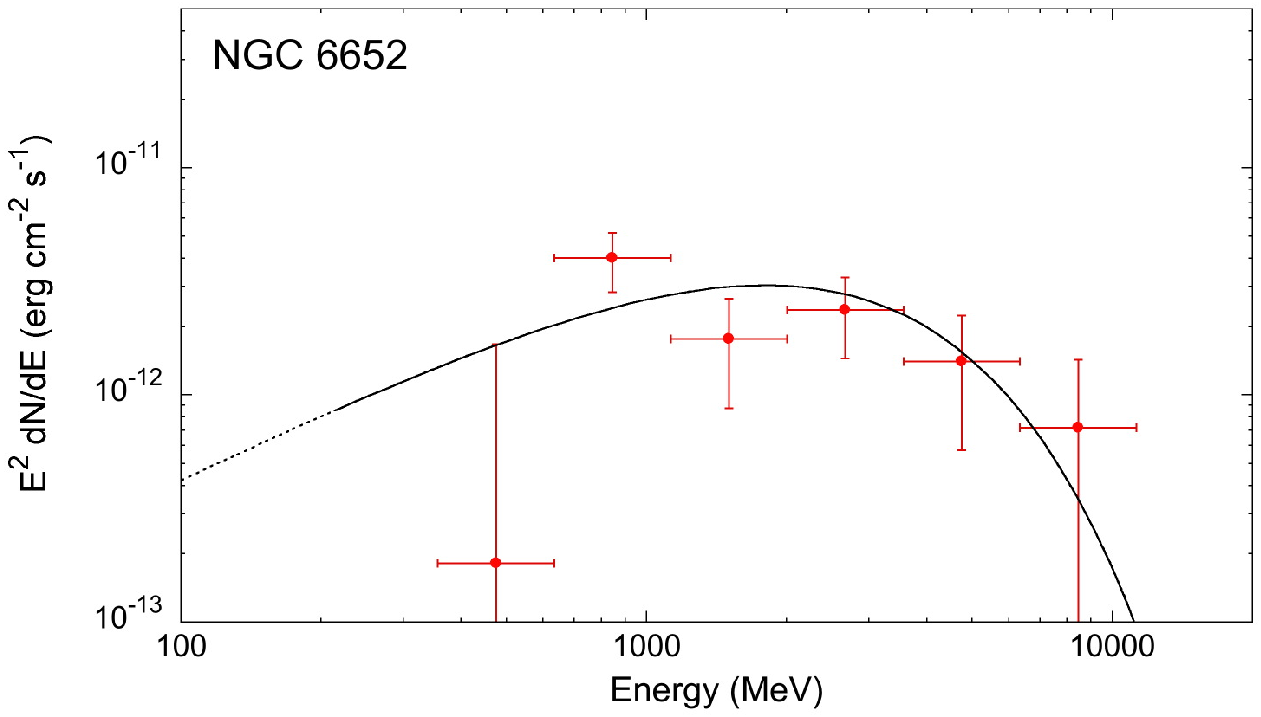}
\caption{
Spectral energy distributions of the 8 significant globular cluster candidates.
Lines indicate the exponentially cut-off power law models that have been fitted to the data
using the binned maximum likelihood fitting procedure, and for which the spectral parameters
are quoted in Table \ref{tab:results}.
The dotted part of the lines indicate the extrapolation of the exponentially cut-off power law
model towards energies below $200$~MeV.
\label{fig:spectra}
}
\end{figure*}

Table \ref{tab:results} quotes also the variability indicator $\chi^2_{\rm month}$, which is inferior
to $26.3$ for all globular cluster candidates, indicating that none of the gamma-ray sources exhibit
significant time variability.
All 8 sources thus match the expected property of globular clusters for being non-variable sources
of gamma rays.
Table \ref{tab:results} also shows that the 8 significant sources exhibit the expected 
spectral characteristics of globular clusters: a flat ($\Gamma<2$) power law spectral index
with an exponential cut-off in the few GeV domain.
For two of the sources (M~62 and NGC~6440), however, the statistical significance of the
cut-off is $\scut<3\sigma$. More data are clearly needed here before definite conclusions about the reality of the exponential
cut-off can be drawn.
Finally, we show in Fig.~\ref{fig:spectra} the spectral energy distributions of the 8 significant
globular cluster candidates.
In the following we comment on the analysis results for each of these globular clusters.

\subsection{Comments on individual clusters}

\subsubsection{47~Tucanae}

47~Tuc was the first globular cluster to be associated with a gamma-ray source
detected by {\it Fermi} LAT \citep{abdo09c}.
The maximum likelihood position of the gamma-ray source is only $1.7'$ offset from the
cluster core, which is well inside the 95\% confidence error radius of $3.3'$.
The spatial and spectral characteristics that we derive from the present dataset are compatible 
with those published in our previous work \citep{abdo09c}, yet the $\sim2.8$ times longer
exposure now allows us to constrain the spectral source parameters considerably better.
In particular, the uncertainty on the energy flux of 47~Tuc has been reduced by a factor 
$\sim2$ to less than $10\%$, improving the constraint on the total gamma-ray luminosity of the
cluster, and hence on the expected population of millisecond pulsars 
(cf.~Section \ref{sec:discussion}).

\subsubsection{Omega Cen}

Omega Cen is one of the globular clusters in our list for which no MSP has so far been
detected.
Omega Cen is situated near the southern giant radio-lobe of the nearby
Cen~A radio galaxy which is also an extended source of GeV gamma rays
\citep{abdo10c}.
To account for emission from the radio lobes in our analysis we included a spatial template
based on the 22 GHz WMAP microwave map \citep{hinshaw09} in our background model.
Formally, the position of the gamma-ray source near Omega Cen (1FGL~J1328.2$-$4729) 
is inconsistent with the location of the globular cluster.
It turned out, however, that 1FGL~J1328.2$-$4729 is a superposition of two distinct sources
that can be separated thanks to their different spectral characteristics.
While the first source shows a cut-off in the few GeV domain, the second source is rather 
hard ($\Gamma=1.6 \pm 0.2$), with no cut-off and can be clearly isolated from the first one by selecting 
only events $>10$~GeV.
This results in a best fitting location
of $\ra=13^{\rm h}28.7^{\rm m}$ and $\dec=-47\deg30'$ with a $95\%$ error radius
of $r_{95}=4.0'$.
Alternatively, we fixed a point source at the location of Omega Cen and fitted the location
of the residual counts to
$\ra=13^{\rm h}28.8^{\rm m}$ and $\dec=-47\deg29'$ with a $95\%$ error radius
of $r_{95}=3.1'$.
Both locations are spatially consistent with the radio source
SUMSS~J132840$-$472748,
which has formally been associated to 1FGL~J1328.2$-$4729 in the 
{\it Fermi} LAT first year catalogue of active galactic nuclei \citep{abdo10d}.
The gamma-ray source is also consistent with the radio source
PMN~J1328$-$4728 and the X-ray sources 1WGA~J1328.6$-$4727 and
1RXS~J132846.7$-$472759, which all are possibly just different designators for
the same source detected by other instruments.

We thus include this source in our background model and re-determine the localisation
of the source that is only visible below a few GeV.
This results in a maximum likelihood position of
$\ra=13^{\rm h}26.5^{\rm m}$ and $\dec=-47\deg29'$ with a $95\%$ error radius
of $r_{95}=7.5'$.
This position is $3.2'$ offset from the centre of  Omega Cen, and thus the gamma-ray 
source is spatially fully consistent with the location of the globular cluster.

\subsubsection{M~62}

M~62 has formally been associated with 1FGL~J1701.1$-$3005 \citep{abdo10b}
and also our maximum likelihood position is only $1.6'$ offset from the core of
the globular cluster, which is thus well within $r_{95}=4.4'$.
The gamma-ray source shows only marginal ($\scut=2.5$) evidence for an 
exponential cut-off and exhibits the largest formal cut-off energy 
($\ecut=4.4^{+3.8+17.7}_{-1.8-1.8}$ GeV) 
of all sources.
The SED (cf.~Fig.~\ref{fig:spectra}) shows no clear indication for a distinct 
cut-off, and if there were not a lack of flux below $\sim600$~MeV,
the SED would probably also be compatible with a soft power law.
M~62 is located in a region of intense Galactic diffuse emission (angular distance
from the Galactic centre $<10\deg$; cf.~Fig.~\ref{fig:images}) which makes any 
proper extraction of spectral parameters difficult.
In particular, we recognised that the source spectrum is rather sensitive to the exact
choice of the ROI, which is readily explained by systematic uncertainties in the
model of Galactic diffuse emission that we used for the analysis.
We thus qualify the gamma-ray source associated with M~62 only as {\it possible} 
globular cluster candidate (in contrast to being a {\it plausible} candidate),
and await the accumulation of more data before drawing any definite conclusions.

\subsubsection{NGC~6388}

NGC 6388 has formally been associated with 1FGL~J1735.9$-$4438 \citep{abdo10b}.
Our position is consistent with that of 1FGL~J1735.9$-$4438 and NGC~6388 is found
right on the 95\% confidence contour of the gamma-ray source, at an angular separation 
of $5.7'$.
The SED shows a clear cut-off at $\ecut=1.8^{+1.2+1.8}_{-0.7-0.6}$ GeV which is typical
of the cut-off energies that are observed for MSPs \citep{abdo09b}.
The {\it Fermi} LAT source thus qualifies as a plausible candidate for a 
globular cluster.
We note that no MSP has so far been detected in this GC.

\subsubsection{Terzan~5}

Terzan~5 has formally been associated with 1FGL~J1747.9$-$2448 \citep{abdo10b}
and our maximum likelihood position is offset by $2.4'$ from the cluster centre,
which is inferior to $r_{95}=2.9'$ for this source.
1FGL~J1747.9$-$2448 has recently been studied by \citet{kong10}.
Using an exponentially cut-off power law model, they obtained 
a spectral index of $\Gamma=1.9\pm0.2$, 
a cut-off energy of $\ecut=3.8\pm1.2$ GeV, and 
$0.1-10$~GeV photon and energy fluxes of
$2 \times 10^{-7}$ \funit\ and 
$1.2 \times 10^{-10}$ \eunit, respectively.
Our analysis confirms the presence of the exponential cut-off at a significance
of $7.1\sigma$, yet with slightly different spectral parameters.
In particular, our fluxes are lower by about a factor of 2, and our SED 
(our Fig.~\ref{fig:spectra})
more closely follows an exponentially cut-off model rather than the SED derived by
\citet{kong10} (their Fig.~2).
Possibly, the choice of a rather large ROI with a radius of $15\deg$ for their 
analysis introduced some systematic uncertainties due to the inaccurate 
subtraction of  Galactic diffuse emission in this complex region near the Galactic 
centre.
We used a rectangular ROI of $6\deg \times 6\deg$ for our analysis which
considerably reduces the impact of the Galactic diffuse model on the analysis,
and which allows for a more reliable determination of the spectral parameters of
the source.
The spectral parameters we obtain for Terzan~5 are indeed very close to those
observed for 47~Tuc, making the gamma-ray source a very plausible candidate
for being a globular cluster.

\subsubsection{NGC~6440}

NGC~6440 has formally been associated with 1FGL~J1748.7$-$2020 \citep{abdo10b},
and our fitted position is offset by $1.3'$ from the cluster core, well within $r_{95}=5.2'$.
The SED, however, shows no convincing evidence for an exponential cut-off, and
also the spectral analysis results in only $\scut=1.4\sigma$, too small to claim the
detection of a cut-off.  The spectral analysis may have suffered from the proximity of the Galactic plane
and the related uncertainties due to the subtraction of the diffuse emission, as well as the proximity of a bright gamma-ray pulsar (PSR~J1741$-$2054; \citet{abdo09d}).
In any case, the gamma-ray source associated with NGC~6440 is a relatively faint
source, and more data are required to allow for an accurate spectral characterisation
of the emission.
Pending the acquisition of these data we thus qualify the source only as a
possible candidate for being a globular cluster.

\subsubsection{M~28}

M~28 has been formally associated with 1FGL~J1824.5$-$2449 
\citep{abdo10b}, and our source position is offset by only $1.6'$ from the centre
of the cluster, well within the 95\% error radius of $8.0'$.
Figure~\ref{fig:images} suggests a somewhat elongated shape for the gamma-ray
source that may indicate a superposition of two objects, yet the source is formally 
consistent with a point source, although with the largest of all upper limits of
$\ext<15.6'$.
The SED of M~28 shows a clear exponential cut-off, which is detected at a
significance level of $4.3\sigma$, the third largest significance after Terzan~5 and
47~Tuc.
The gamma-ray source associated to M~28 thus qualifies as a plausible globular
cluster candidate.

Note that \citet{pellizzoni09} have reported the detection (at $4.2\sigma$ significance)
of pulsations from PSR~J1824$-$2452A with the {\it AGILE} satellite during a 
5-day interval in August 2007, while no pulsations have been seen for other periods
of comparable duration.
PSR~J1824$-$2452A is one of the 12 known MSPs in M~28 and ranks among the
youngest and most energetic millisecond pulsars that are known
\citep{knight06}.
The $>$100~MeV flux of $(18 \pm 5) \times 10^{-8}$ \funit\ that \citet{pellizzoni09} quote
for the 5-day period is significantly larger than the flux of $(2.6 \pm 1.3) \times 10^{-8}$ 
\funit\ that we measured from the cluster using our dataset.
Searches for pulsations from PSR~J1824$-$2452A in the {\it Fermi} LAT data have so
far not revealed any significant detection.
Furthermore, we did not find any evidence for significant flux variations from M~28 on 
time-scales from 1 week to 1 month.
Typical flux measurement uncertainties for 1 week, however, amount to
$\pm8 \times 10^{-8}$ \funit\ ($1\sigma$),
hence flares from M~28 of the duration and amplitude quoted by {\it AGILE} formally 
cannot be excluded by our data.

\subsubsection{NGC~6652}

NGC~6652 has been formally associated with 1FGL~J1835.3$-$3255
\citep{abdo10b}, and our source position is offset by $1.7'$ from the
cluster, well below $r_{95}=7.5'$.
The SED does not reveal a convincing exponential cut-off although the
spectral analysis suggests $\scut=3.2\sigma$, mainly due to the lack of
flux below $\sim600$ MeV.
The gamma-ray source associated with NGC~6652 is a rather faint source,
and also here, more data are required to better constrain the SED before
definite conclusions should be drawn.
Pending the acquisition of these data we thus qualify the source only as a
possible candidate for being a globular cluster.
Establishing a firm exponential cut-off is particularily interesting in this case
since so far no MSP has been detected in NGC~6652.

\section{Discussion}
\label{sec:discussion}

\subsection{Omega Cen, NGC~6388 and NGC~6652}

Three of the globular clusters for which we detected gamma-ray
counterparts, Omega Cen, NGC~6388 and NGC~6652, deserve
attention, since so far no MSPs have been detected in these
objects.
Present pulsar searches in the radio and X-ray domains are severely
limited by the instrument sensitivity, and only in a few examples
(e.g.~47~Tuc and M~15) has the end of the pulsar luminosity function possibly 
been reached \citep{camilo05}.
In some cases, this may simply be because pulsars are generally intrinsically
weak objects, and globular clusters are often distant.
Interstellar scattering, which broadens pulsations because of multipath
propagation, can also be a major obstacle, especially for MSPs in clusters
at low Galactic latitudes.
Furthermore, interstellar absorption hinders detection in the X-ray range, in particular
for clusters that are situated near the Galactic plane.

NGC~6388 and NGC~6652 are indeed the most distant objects in
our list of globular clusters for which gamma-ray emission is significantly 
detected (see Table~\ref{tab:derived}).
Among the 26 globular clusters in which MSPs have so far been
detected\footnote{See
http://www.naic.edu/$\sim$pfreire/GCpsr.html},
6 have distances that are comparable to or larger than those of
NGC~6388 and NGC~6652.
Four of these clusters (NGC~1851, M~53, M~3, and M~15) show
considerably lower dispersion measures than NGC~6388 and NGC~6652,
hence radio pulsars are more easily detected in these systems despite
their large distances.
The two other clusters (NGC~6441 and NGC~6517) have dispersion measures
that are comparable to the estimates for NGC~6388 and NGC~6652, 
yet uncertainties in the dispersion measure estimates \citep{cordes02}
or selection effects in the radio surveys that are very difficult to quantify
\citep{hessels07} may explain why no MSPs have so far been detected in 
these clusters.

Omega Cen is substantially closer than NGC~6388 and NGC~6652, at
a distance that is comparable to that of 47~Tuc \citep{bono08}.
However, while 47~Tuc is situated at high galactic latitudes with
a correspondingly small dispersion measure of 
$\sim25$~pc cm$^{-3}$,
Omega Cen is closer to the galactic plane and has an estimated
dispersion measure of $126$~pc cm$^{-3}$ \citep{cordes02}, so
MSPs searches will be more difficult in Omega Cen compared to
47~Tuc.
Omega Cen has been searched for radio pulsars by \cite{edwards01}
with the Parkes 64~m radio telescope at 660 MHz without success.
Current X-ray observations of this cluster with {\em Chandra} and 
{\em XMM-Newton} \citep{hagg09,gend03b} are not deep enough to
uniquely identify MSPs. 
However, \cite{hagg09} note that their Chandra observations do not rule
out the presence of MSPs, as numerous sources occupy the region of the 
colour-magnitude diagram (CMD) where MSPs would be expected to lie.
They compared their X-ray CMD with that of \cite{grin01} for 47 Tuc and
adjusted for the somewhat larger hydrogen column density to Omega Cen 
versus 47 Tuc (and slightly different distances and exposure times).
Eighteen objects are found in the range of brightness and colour expected
for MSPs in Omega Cen.
This range of colours is also typical of chromospherically active single
stars and binaries, but many or all of these objects may be MSPs.
Our estimate is also commensurate with the 28 (with an rms dispersion of 6)
MSPs predicted via two body tidal capture \citep{dist92}.

\subsection{Expected MSP populations}

\begin{table}[!t]
\footnotesize
\caption{Isotropic gamma-ray luminosities and expected numbers of MSPs.
\label{tab:derived}
}
\begin{center}
\begin{tabular}{lrcc}
\hline
\hline
\noalign{\smallskip}
Name & \multicolumn{1}{c}{$d$ (kpc)} & \multicolumn{1}{c}{\liso $(10^{34} \ergs)$} &
   \multicolumn{1}{c}{\nmsp} \\
\noalign{\smallskip}
\hline
\noalign{\smallskip}
47 Tucanae & $4.0 \pm 0.4^{\,(1)}$ & $4.8^{+1.1}_{-1.1}$ & $33^{+15}_{-15}$ \\
\noalign{\smallskip}
Omega Cen & $4.8 \pm 0.3^{\,(2)}$ & $2.8^{+0.7}_{-0.7}$ & $19^{+9}_{-9}$ \\
\noalign{\smallskip}
M~62            & $6.6 \pm 0.5^{\, (3)}$ & $10.9^{+3.5}_{-2.3}$ & $76^{+38}_{-34}$ \\
\noalign{\smallskip}
NGC~6388  & $11.6 \pm 2.0^{\,(4)}$ & $25.8^{+14.0}_{-10.6}$ & $180^{+120}_{-100}$ \\
\noalign{\smallskip}
Terzan~5     & $5.5 \pm 0.9^{\,(5)}$ & $25.7^{+9.4}_{-8.8}$ & $180^{+100}_{-90}$ \\
\noalign{\smallskip}
NGC~6440  & $8.5 \pm 0.4^{\, (6)}$ & $19.0^{+13.1}_{-5.0}$ & $130^{+100}_{-60}$ \\
\noalign{\smallskip}
M~28            & $5.1 \pm 0.5^{\, (7)}$ & $6.2^{+2.6}_{-1.8}$ & $43^{+24}_{-21}$ \\
\noalign{\smallskip}
NGC~6652  & $9.0 \pm 0.9^{\, (8)}$ & $7.8^{+2.5}_{-2.1}$ & $54^{+27}_{-25}$ \\
\noalign{\smallskip}
\hline
\noalign{\smallskip}
NGC~6541  & $6.9 \pm 0.7^{\, (9)}$ & $<4.7$ & $<47$ \\
NGC~6752  & $4.4 \pm 0.1^{\, (10)}$ & $<1.1$ & $<11$ \\
M~15            & $10.3 \pm 0.4^{\,(11)}$ & $<5.8$ & $<56$ \\
\noalign{\smallskip}
\hline
\end{tabular}
\end{center}
{\bf Notes.}
References to the distance estimates are:
(1) \citet{mclaughlin06};
(2) \citet{vandeven06};
(3) \citet{beccari06};
(4) \citet{moretti09};
(5) \citet{ortolani07};
(6) \citet{ortolani94};
(7) \citet{rees91};
(8) \citet{chaboyer00};
(9) \citet{lee06};
(10) \citet{sarajedini07};
(11) \citet{vandenbosch06}.
\end{table}

Under the assumption that MSPs share the same characteristics in the various globular clusters
that we studied, we may use the observed gamma-ray luminosities to obtain estimates on the total
number of MSPs expected to reside in each cluster.
Here we closely follow the formalism that has been developed by \citet{abdo09c} to estimate
the size of the MSP population in 47~Tuc.
The total number \nmsp\ of MSPs present in a globular cluster is estimated
from the isotropic gamma-ray luminosity of the cluster \liso,
the average spin-down power \edot\ of MSPs, and
the estimated average spin-down to gamma-ray luminosity 
conversion efficiency \eff\ using
\begin{equation}
\nmsp = \frac{\liso}{\edot \eff} .
\label{eq:nmsp}
\end{equation}
We compute the isotropic gamma-ray luminosity using
$\liso = 4 \pi S d^2$,
where $S$ is the observed energy flux and $d$ the distance to the globular cluster.
$\edot$ 
for the pulsar populations in individual clusters are only poorly known
since measurements of 
$\dot{E}$ are significantly affected by acceleration in the gravitational
potential of the globular cluster \citep{bogdanov06}.
We therefore adopt a value of $\edot = (1.8 \pm 0.7) \times 10^{34}$ \ergs\
for all clusters as a best estimate that has been derived by comparing
the $\log \dot{P}$ distribution of Galactic field MSPs to the acceleration 
corrected $\log \dot{P}$ distribution obtained for MSPs in 47~Tuc \citep{abdo09c}.
This value is close to the value of 
$\edot = 3.0 \times 10^{34}$ \ergs\
that is obtained as average for Galactic field MSPs from the ATNF catalogue.
Since intrinsically luminuous pulsars are more easily detected, this average
is likely biased towards high values.
Limiting the average to the local ($<1$~kpc) Galactic field MSPs in the ATNF
catalogue indeed reduces the average value to 
$\edot = 1.1 \times 10^{34}$ \ergs, consistent with our adopted value.
Furthermore, we assume $\eff = 0.08$ as derived from observations of
nearby MSPs \citep{abdo09c}.
The results are summarised in Table~\ref{tab:derived}.
The statistical and systematic uncertainties in the energy flux have been
added in quadrature and we quote the resulting uncertainty from that
combination for \liso\ and \nmsp.

It is encouraging to note that the estimates of MSP populations in the
above globular clusters are very similar to theoretical and/or
observational constraints made in other wavelengths.
In Terzan 5, 33 MSPs are known so far and many more are expected
based on the extrapolation of the radio luminosity function
\citep{ransom05}.
Indeed, by comparing the total radio luminosity of Terzan~5
to the radio luminosity function of field pulsars, \citet{fruchter00}
estimate the number of pulsars in the cluster to be $60-200$, and also
\citet{kong10} estimate that the number of pulsars in Terzan~5 could
be as large as $\sim200$.
For NGC~6440, \cite{freire08} estimate that it may house as many pulsars as 
Terzan 5, a prediction that is compatible with our observations.
\cite{hein05} estimate that a total of $\sim$25 MSPs exist in 47 Tuc
($<$60 at 95\% confidence), regardless of their radio beaming fraction,
based on their deep {\em Chandra} X-ray observations of this cluster.
This is comparable to the upper limit of 30 MSPs derived by 
\cite{mcconnell04} from the radio luminosity distribution of pulsars.
As noted above, \cite{hagg09} and \cite{dist92} obtain similar estimates
for the number of MSPs in Omega Cen as well.
\cite{kulk90} used radio observations of Galactic globular clusters visible
from the northern hemisphere and took into account selection effects in
the various radio surveys, and then after estimating the relative efficiency
of pulsar production in the individual clusters, produced a census of the
cluster population of pulsars.
Their estimates are surprisingly close to ours where they deduce 150 MSPs
in Terzan 5, 86 in NGC~6440, 18 in M~28, 21 in M~15 but only 5 in 
NGC~6652, the latter being the only cluster for which they estimate a
quantity that is remarkably dissimilar to our work 
(see Table~\ref{tab:derived}).  
We recall, however, that we classified NGC~6652 only as a {\it possible} 
globular cluster candidate because of the rather weak evidence for
an exponential cut-off in the spectrum.
The acquisition of more data is definitely required before any firm conclusions
about the number of MSPs in NGC~6652 should be drawn.

\subsection{Stellar encounters and the formation of MSPs}

It has been shown for globular clusters that the number of neutron star X-ray 
binaries is correlated with the stellar encounter rate
\erate\ \citep[see Section 1 and][]{gendre03}.  Since MSPs are believed to be the progeny
of these systems, we may expect a similar  correlation between our
estimated number of MSPs (or equivalently the observed gamma-ray
luminosity) and \erate. 

To test the link between the formation of MSPs and the stellar
encounter rate, we compute the latter using $\erate = \rho_0^{1.5}
r_{\rm c}^2$, where $\rho_0$ is the central cluster density, given in
units of \Lsol\ pc$^{-3}$, and $r_{\rm c}$ is the cluster core radius,
given in units of pc.  We used the central cluster densities and
angular core radii $\theta_c$ quoted in  \citet{harris96} (February
2003 revision) along with the distance estimates listed in
Table~\ref{tab:derived} using $\rho_0 = \rho_0^{\rm Harris} d^{\rm
  Harris} / d$ \citep{djorgovski93}, and using $r_{\rm c} = d \tan
\theta_c$.  We then normalised the stellar encounter rates so that
$\erate=100$ for M~62 (for M~62 we obtain 
$\rho_0^{1.5} r_{\rm c}^2=6.5 \times 10^6$~\Lsol$^{1.5}$~pc$^{-2.5}$).

\begin{figure}[!t]
\centering \includegraphics[width=9cm]{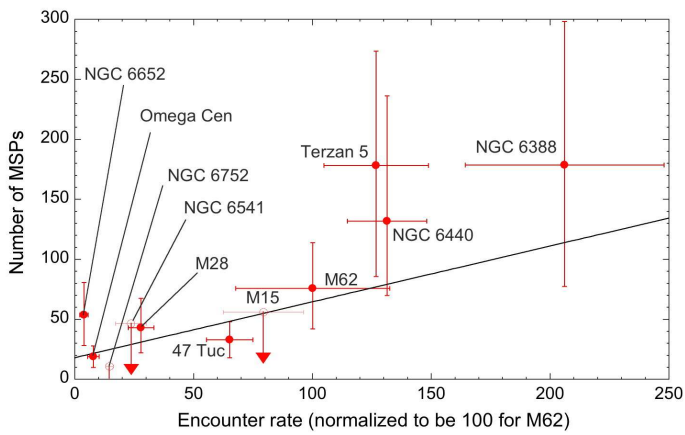}
\caption{Predicted number of MSPs versus stellar encounter rate \erate.
  Horizontal error bars indicate the uncertainties in \erate\ due to the distance
  uncertainties given in Table \ref{tab:derived} and due to uncertainties in
  $\theta_{\rm c}$ that we estimated from the spread of values quoted in the recent
  literature.
  The data have been fitted by a linear relation $\nmsp = 0.5 \times \erate + 18$.
\label{fig:encounter}
}
\end{figure}

Figure~\ref{fig:encounter} shows evidence for a linear correlation
between the predicted numbers of MSPs and the stellar encounter rates.
Taking into account the uncertainties, the correlation is well fitted
by 
$\nmsp = (0.5 \pm 0.2) \times \erate + (18 \pm 9)$
(correlation coefficient $0.7$).
A similar
correlation (not shown) is obtained if we plot the isotropic gamma-ray
luminosity  versus the encounter rate, since \nmsp\ and \liso\ are
linearly related by Eq.~(\ref{eq:nmsp}).  The correlation is well
fitted by 
$\liso = (5.5 \pm 1.9) \times 10^{32} \times \erate + (2.6 \pm 0.7) \times 10^{34}$~\ergs.

Using our correlation we predict 
$2600-4700$ 
MSPs that are observable in gamma rays in Galactic globular clusters.
\cite{kulk90} estimate a total of $\sim10,000$ pulsars in Galactic globular
clusters, subject to uncertainties in pulsar beaming and binarity.
It has been suggested that the rare electron-capture accretion induced collapse
channel will have a low kick velocity and may thus dominate the final cluster MSP
population \citep{ivan08}.
However, since binaries dominate the production channels, the extrapolation 
from the observed sample to the total population based on encounter rate,
used for the estimate of the total pulsar population by \cite{kulk90},
should still be relevant.
On the low end, \cite{hein05} estimated the Galactic globular cluster
pulsar population as 700, essentially from X-ray observations
and the stellar encounter rate, the latter being commensurate with the
estimate of \cite{wije91} which was deduced from radio observations.
\cite{fruchter90} deduced the total number of MSPs in the
Galactic globular system to lie between 500 and 2000, but
concluded that the expected number of pulsars in a globular cluster
depends only weakly on the stellar collision rate. 

It appears that our independent method of determining the number of MSPs in
Galactic globular clusters through gamma-ray observations is entirely
compatible with earlier estimates.

\section{Conclusions}
\label{sec:conclusion}

An analysis of {\it Fermi} LAT data from 13 globular clusters has revealed 8 significant, point-like
and steady gamma-ray sources that are spatially consistent with the locations of the clusters.
Five of them (47~Tuc, Omega Cen, NGC~6388, Terzan~5, and M~28) show hard spectral
power law indices $(0.7 < \Gamma < 1.4)$ and clear evidence for an exponential cut-off in the 
range $1.0-2.6$ GeV, which is the characteristic signature of magnetospheric emission from 
MSPs.
We thus classify these 5 sources as {\it plausible} globular cluster candidates.
Three of them (M~62, NGC~6440 and NGC~6652) also show hard spectral indices
$(1.0 < \Gamma < 1.7)$, however the presence of an exponential cut-off cannot 
unambiguously be established.
More data are required before definite conclusions can be drawn; hence we qualify these
3 sources as {\it possible} globular cluster candidates.

From the 8 globular clusters that are associated with significant gamma-ray sources, 5 are 
known to harbour MSPs.
In Omega Cen, NGC~6388 and NGC~6652, however, no MSPs have so far been detected,
neither by radio nor by X-ray observations.
The observation of gamma-ray signatures that are characteristic of MSPs provides strong support that these GCs indeed also harbour important populations
of MSPs.
In particular, we predict from the observed gamma-ray luminosities that the total MSP 
populations amount to
$10-30$ (Omega Cen),
$80-300$ (NGC~6388), and
$30-80$ (NGC~6652) in these clusters.
Deep radio and X-ray follow-up observations may help to unveil first members of these
populations.

Our predicted number of MSPs shows evidence for a positive correlation with the stellar 
encounter rate in a similar way to their progenitors, the neutron star low mass X-ray binaries.
This correlation allows us to deduce the total number of MSPs in Galactic globular clusters 
($2600-4700$) 
which lies midway between all previous estimates, supporting such a correlation.  
Such an estimate can be used to derive constraints on the original neutron star
X-ray binary population, essential for understanding the importance of binary
systems in slowing the inevitable core collapse of globular clusters.

\begin{acknowledgements}
The \textit{Fermi} LAT Collaboration acknowledges generous ongoing support
from a number of agencies and institutes that have supported both the
development and the operation of the LAT as well as scientific data analysis.
These include the National Aeronautics and Space Administration and the 
Department of Energy in the United States, the Commissariat \`a l'Energie Atomique
and the Centre National de la Recherche Scientifique / Institut National de Physique
Nucl\'eaire et de Physique des Particules in France, the Agenzia Spaziale Italiana
and the Istituto Nazionale di Fisica Nucleare in Italy, the Ministry of Education,
Culture, Sports, Science and Technology (MEXT), High Energy Accelerator Research
Organization (KEK) and Japan Aerospace Exploration Agency (JAXA) in Japan, and
the K.~A.~Wallenberg Foundation, the Swedish Research Council and the
Swedish National Space Board in Sweden.

Additional support for science analysis during the operations phase is gratefully
acknowledged from the Istituto Nazionale di Astrofisica in Italy and the 
Centre National d'\'Etudes Spatiales in France.
\end{acknowledgements}


\end{document}